\documentclass[graybox]{svmult}

\usepackage{helvet}         
\usepackage{courier}        
\usepackage{type1cm}        
\usepackage{makeidx}         
\usepackage{graphicx}        
\usepackage{multicol}        
\usepackage[bottom]{footmisc}
\usepackage{amssymb,latexsym,amsmath,amsbsy}
\usepackage{cite}
\usepackage{stmaryrd}  
\usepackage{arydshln}  
\usepackage{xcolor}

\def\nn{\nonumber}

\def\qdots{\mathinner{\mkern1mu\raise1pt\vbox{\kern7pt\hbox{.}}\mkern2mu \raise4pt\hbox{.}\mkern2mu\raise7pt\hbox{.}\mkern1mu}}
\def\Z{{\mathbb Z}}

\def\osp{\mathfrak{osp}}
\def\pso{\mathfrak{pso}}

\begin{document}

\title*{A Klein operator for paraparticles}
\author{N.I.~Stoilova and J.\ Van der Jeugt}
\institute{N.I.~Stoilova \at Institute for Nuclear Research and Nuclear Energy,
Bulgarian Academy of Sciences,  
Boul.\ Tsarigradsko Chaussee 72, 1784 Sofia, Bulgaria, \email{stoilova@inrne.bas.bg}
\and J.\ Van der Jeugt \at Department of Applied Mathematics, Computer Science and Statistics, Ghent University,
Krijgslaan 281-S9, B-9000 Gent, Belgium \email{Joris.VanderJeugt@UGent.be}}
\maketitle

\abstract{It has been known for a long time that there are two non-trivial possibilities for the 
relative commutation relations between a set of $m$ parafermions and a set of
$n$ parabosons.
These two choices are known as ``relative parafermion type'' and ``relative paraboson type'', 
and correspond to quite different underlying algebraic structures.
In this short note we show how the two possibilities are related by a so-called Klein transformation.}

\vskip 1cm
\setcounter{equation}{0}

The standard creation and annihilation operators of identical particles satisfy canonical commutation (boson) or
anticommutation (fermion) relations, expressed by means of commutators or anticommutators.
In 1953 Green~\cite{Green} generalized bosons to so-called parabosons and fermions to parafermions, by postulating
certain triple relations for the creation and annihilation operators, rather than just (anti)commutators.
A system of $m$ parafermion creation and annihilation operators $f_j^\pm$ ($j=1,\ldots,m$) is determined by
\begin{equation}
[[f_{ j}^{\xi}, f_{ k}^{\eta}], f_{l}^{\epsilon}]=
|\epsilon -\eta| \delta_{kl} f_{j}^{\xi} - |\epsilon -\xi| \delta_{jl}f_{k}^{\eta}, 
\label{f-rels}
\end{equation}
where $j,k,l\in \{1,2,\ldots,m\}$ and $\eta, \epsilon, \xi \in\{+,-\}$ (to be interpreted as $+1$ and $-1$
in the algebraic expressions $\epsilon -\xi$ and $\epsilon -\eta$).
Similarly, a system of $n$ pairs of parabosons $b_j^\pm$ satisfies
\begin{equation}
[\{ b_{ j}^{\xi}, b_{ k}^{\eta}\} , b_{l}^{\epsilon}]= 
(\epsilon -\xi) \delta_{jl} b_{k}^{\eta}  + (\epsilon -\eta) \delta_{kl}b_{j}^{\xi}.
\label{b-rels}
\end{equation}
These triple relations involve nested (anti)commutators, just like the Jacobi identity of Lie (super)algebras.
And indeed, later it was shown~\cite{Kamefuchi,Ryan} that the parafermionic algebra determined by~\eqref{f-rels} is 
the orthogonal Lie algebra $\mathfrak{so}(2m+1)$,
and that the parabosonic algebra determined by~\eqref{b-rels} is the orthosymplectic Lie superalgebra 
$\mathfrak{osp}(1|2n)$~\cite{Ganchev}. 

Greenberg and Messiah~\cite{GM} considered combined systems of parafermions and parabosons. 
Apart from two trivial combinations (where the parafermions and parabosons mutually commute or anticommute), 
they found two non-trivial relative commutation relations between parafermions and parabosons, also expressed by
means of triple relations. 
The first of these are the relative parafermion relations, determined by:
\begin{align}
&[[f_{ j}^{\xi}, f_{ k}^{\eta}], b_{l}^{\epsilon}]=0,\qquad [\{b_{ j}^{\xi}, b_{ k}^{\eta}\}, f_{l}^{\epsilon}]=0, \nn\\
&[[f_{ j}^{\xi}, b_{ k}^{\eta}], f_{l}^{\epsilon}]= -|\epsilon-\xi| \delta_{jl} b_k^{\eta}, \qquad
\{[f_{ j}^{\xi}, b_{ k}^{\eta}], b_{l}^{\epsilon}\}= (\epsilon-\eta) \delta_{kl} f_j^{\xi}.
\label{rel-pf}
\end{align}
The second are the so-called relative paraboson relations, and will appear later in this paper.

The parastatistics algebra with relative parafermion relations, determined by~\eqref{f-rels}, 
\eqref{b-rels} and~\eqref{rel-pf}, was identified by
Palev~\cite{Palev1} and is the Lie superalgebra $\mathfrak{osp}(2m+1|2n)$.

When dealing with parastatistics, a major object of study is the Fock space.
By definition the parastatistics Fock space of order $p$ is the Hilbert space with vacuum vector $|0\rangle$, 
defined by means of 
\begin{align}
& \langle 0|0\rangle=1, \qquad f_j^- |0\rangle = 0, \qquad b_j^- |0\rangle = 0,
\qquad (f_j^\pm)^\dagger = f_j^\mp, \qquad (b_j^\pm)^\dagger = b_j^\mp,\nn\\
& [ f_j^-,f_k^+ ] |0\rangle = p\delta_{jk}\, |0\rangle, 
\qquad \{ b_j^-,b_k^+ \} |0\rangle = p\delta_{jk}\, |0\rangle,
\label{Fock}
\end{align}
and by irreducibility under the action of the elements $f_j^\pm$, $b_j^\pm$. 

The purpose of this short contribution is to show that a Klein operator~\cite{Klein,Luders,Vasiliev1984} 
can be constructed, and that new operators $\tilde b_k^{\pm}$ and $\tilde f_j^{\pm}$ can be defined in terms of the operators $b_k^{\pm}$ and $f_j^{\pm}$ in such a way that these new operators satisfy the parastatistics algebra with relative paraboson relations.

First of all, let us define the following elements in terms of the paraoperators $b_j^{\pm}$ and $f_i^{\pm}$
\begin{align}
& h_i=-\frac12[ f_i^-, f_i^+] \qquad (i=1,\ldots,m) \nn\\
& h_{m+j}=\frac12\{  b_j^-, b_j^+\} \qquad (j=1,\ldots,n). 
\label{h-basis}
\end{align}
From the triple relations~\eqref{f-rels}, \eqref{b-rels} and~\eqref{rel-pf}, it is easy to deduce that
\begin{align}
& h_i f_j^\pm=f_j^\pm (h_i\pm \delta_{ij}) \qquad(i=1,\ldots,m;\; j=1,\ldots,m) \nn\\
& h_i b_k^\pm=b_k^\pm h_i \qquad(i=1,\ldots,m;\; k=1,\ldots,n) \nn\\
& h_{m+i} f_j^\pm=f_j^\pm h_{m+i} \qquad(i=1,\ldots,n;\; j=1,\ldots,m) \nn\\
& h_{m+i} b_k^\pm=b_k^\pm (h_{m+i}\pm \delta_{ik}) \qquad(i=1,\ldots,n;\; k=1,\ldots,n).
\label{h-rels}
\end{align}
So if we put
\begin{equation}
H=h_1+h_2+\cdots +h_{m+n},
\label{H}
\end{equation}
then
\begin{align}
& H f_j^\pm = f_j^\pm (H\pm 1) \qquad (j=1,\ldots,m)\nn\\
& H b_k^\pm = b_k^\pm (H\pm 1) \qquad (k=1,\ldots,n).
\label{H-rels}
\end{align}
All these relations are purely algebraic, i.e.\ they follow from the triple relations~\eqref{f-rels}, \eqref{b-rels} and~\eqref{rel-pf}, 
and hold in the algebra generated by the $2(m+n)$ elements $b_k^{\pm}$ and $f_j^{\pm}$.

Following~\eqref{H-rels}, we would like to define an operator of the form $(-1)^H$. 
In order to have a proper meaning for this, it is useful to work in the Fock space of order $p$, characterized by~\eqref{Fock}, 
so that the abstract algebraic elements become operators acting in this Fock space.
Following~\eqref{h-basis}, one finds that
\begin{equation}
H |0\rangle = -\frac{p}{2}(m-n) |0\rangle.
\end{equation}
So it is convenient to define
\begin{equation}
N=H+\frac{p}{2}(m-n)
\end{equation}
and the Klein operator $K$ by
\begin{equation}
K=(-1)^{N}.
\end{equation}
Then $K|0\rangle=|0\rangle$ and $K^2 |0\rangle = |0\rangle$. Since the commutator of $N$ with $f_j^\pm$ and $b_k^\pm$ is the same as that of $H$ (given by~\eqref{H-rels}), 
this implies that $K^2$ acts as the identity operator in the Fock space. 
Moreover, from~\eqref{H-rels} it follows that
\begin{equation}
K f_j^\pm+ f_j^\pm K=0, \qquad K b_k^\pm +b_k^\pm K =0 \qquad (j=1,\ldots,m;\; k=1,\ldots,n).
\end{equation}
In fact, one could also continue to work purely algebraically, and extend the algebra generated by the generators $b_k^{\pm}$ and $f_j^{\pm}$, 
subject to the relations~\eqref{f-rels}, \eqref{b-rels} and~\eqref{rel-pf}, by an abstract element $K$ satisfying
\begin{equation}
K^2=1, \qquad \{K, f_j^\pm\}=0, \qquad \{K, b_k^\pm\} =0 \qquad (j=1,\ldots,m;\; k=1,\ldots,n).
\label{K-rels}
\end{equation}
The previous analysis just shows that such an operator $K$ exists in the Fock space of the paraoperators.

Let us now proceed to the main construction. Define new operators
\begin{align}
& \tilde f_j^\pm = \pm f_j^\pm K = \mp K f_j^\pm \qquad (j=1,\ldots,m)\nn\\
& \tilde b_k^\pm = b_k^\pm \qquad (k=1,\ldots,n).
\label{tilde-fb}
\end{align}
The purpose is now to examine the triple relations for the new set of operators $\tilde b_k^{\pm}$ and $\tilde f_j^{\pm}$.
Since the $\tilde b_k^{\pm}$ are the same as the $b_k^{\pm}$, one has
\begin{equation}
[\{ \tilde b_{ j}^{\xi}, \tilde b_{ k}^{\eta}\} , \tilde b_{l}^{\epsilon}]= 
(\epsilon -\xi) \delta_{jl} \tilde b_{k}^{\eta}  + (\epsilon -\eta) \delta_{kl} \tilde b_{j}^{\xi}.
\label{tb-rels}
\end{equation}
Next, using $\tilde f_{j}^{\xi}=\xi f_j^\xi K$, $\tilde f_{k}^{\eta}=\eta f_k^\eta K$, $K^2=1$ and~\eqref{K-rels}, one has
\[
[\tilde f_{ j}^{\xi}, \tilde f_{ k}^{\eta}] = -\xi\eta [f_{ j}^{\xi}, f_{ k}^{\eta}].
\]
This implies that
\begin{align*}
[[\tilde f_{ j}^{\xi}, \tilde f_{ k}^{\eta}], \tilde f_{l}^{\epsilon}]& =
-\xi\eta\epsilon (|\epsilon -\eta| \delta_{kl} f_{j}^{\xi} - |\epsilon -\xi| \delta_{jl}f_{k}^{\eta}) K\\
&= -\eta\epsilon |\epsilon -\eta| \delta_{kl} \tilde f_{j}^{\xi} +\xi\epsilon |\epsilon -\xi| \delta_{jl} \tilde f_{k}^{\eta}.
\end{align*}
But for the allowed values of $\xi, \eta, \epsilon \in\{-1,+1\}$, one has $-\eta\epsilon |\epsilon -\eta|= |\epsilon -\eta|$, and similar for the second factor above.
Therefore, the elements $\tilde f_j^\pm$ satisfy the usual parafermion triple relations
\begin{equation}
[[\tilde f_{ j}^{\xi}, \tilde f_{ k}^{\eta}], \tilde f_{l}^{\epsilon}]=
|\epsilon -\eta| \delta_{kl} \tilde f_{j}^{\xi} - |\epsilon -\xi| \delta_{jl}\tilde f_{k}^{\eta}.
\label{tf-rels}
\end{equation}
Next, let us turn to the ``relative relations.''
From the earlier observations, it follows already that
\[
[[\tilde f_{ j}^{\xi}, \tilde f_{ k}^{\eta}], \tilde b_{l}^{\epsilon}]=0,\qquad [\{\tilde b_{ j}^{\xi}, \tilde b_{ k}^{\eta}\}, \tilde f_{l}^{\epsilon}]=0.
\]
Furthermore,
\[
\{\tilde f_{ j}^{\xi}, \tilde b_{ k}^{\eta} \} =
\xi f_j^\xi K b_k^\eta + \xi b_k^\eta f_j^\xi K  = -\xi [f_{ j}^{\xi}, b_{ k}^{\eta}] K,
\]
and then
\begin{align*}
\{\{\tilde f_{ j}^{\xi}, \tilde b_{ k}^{\eta}\}, \tilde f_{l}^{\epsilon}\} 
&=-\xi [f_j^\xi,b_k^\eta]K \epsilon f_l^\epsilon K -\xi \epsilon f_l^\epsilon K [f_j^\xi,b_k^\eta]K\\
&=\epsilon\xi [f_j^\xi,b_k^\eta]f_l^\epsilon - \epsilon\xi f_l^\epsilon [f_j^\xi,b_k^\eta] \\
&=\epsilon\xi [[f_{ j}^{\xi}, b_{ k}^{\eta}], f_{l}^{\epsilon}] = -\epsilon\xi |\epsilon-\xi| \delta_{jl} b_k^\eta = |\epsilon-\xi| \delta_{jl} \tilde b_k^\eta.
\end{align*}
In a similar way, one finds
\begin{align*}
[ \{\tilde f_{ j}^{\xi}, \tilde b_{ k}^{\eta}\}, \tilde b_{l}^{\epsilon} ]
&=-\xi [f_j^\xi,b_k^\eta]K b_l^\epsilon  +\xi b_l^\epsilon  [f_j^\xi,b_k^\eta]K\\
&=\xi [f_j^\xi,b_k^\eta]b_l^\epsilon K + \xi b_l^\epsilon [f_j^\xi,b_k^\eta]K \\
&=\xi \{[f_{ j}^{\xi}, b_{ k}^{\eta}], b_{l}^{\epsilon}\} K = \xi(\epsilon-\eta) \delta_{kl} f_j^\xi K = (\epsilon-\eta) \delta_{kl} \tilde f_j^\xi.
\end{align*}
In other words, the new operators $\tilde b_k^{\pm}$ and $\tilde f_j^{\pm}$ satisfy~\eqref{tb-rels}, \eqref{tf-rels} and
\begin{align}
&[[\tilde f_{ j}^{\xi}, \tilde f_{ k}^{\eta}], \tilde b_{l}^{\epsilon}]=0,\qquad
 [\{\tilde b_{ j}^{\xi}, \tilde b_{ k}^{\eta}\}, \tilde f_{l}^{\epsilon}]=0, \nn\\
&\{\{\tilde f_{ j}^{\xi}, \tilde b_{ k}^{\eta}\}, \tilde f_{l}^{\epsilon}\}= 
|\epsilon-\xi| \delta_{jl} \tilde b_k^{\eta}, \qquad
[\{\tilde f_{ j}^{\xi}, \tilde b_{ k}^{\eta}\}, \tilde b_{l}^{\epsilon}]= (\epsilon-\eta) \delta_{kl} \tilde f_j^{\xi}.
\label{rel-pb}
\end{align}
But\eqref{tb-rels}, \eqref{tf-rels} and \eqref{rel-pb} are exactly the relations for a mixed set of paraparticles 
satisfying the relative paraboson relations~\cite{GM, KA}.

In short, we have shown that the simple Klein transformation~\eqref{tilde-fb} maps the paraoperators with relative parafermion relations to the paraoperators with relative paraboson relations.

Observe that the algebra generated by the paraoperators $\tilde b_k^{\pm}$ and $\tilde f_j^{\pm}$, subject to the relations~\eqref{tb-rels}, \eqref{tf-rels} and~\eqref{rel-pb}, 
is no longer (the enveloping algebra of) a Lie algebra or a Lie superalgebra.
It was identified as a certain $\Z_2\times\Z_2$-graded Lie superalgebra in~\cite{YJ,YJ2,KA,Tolstoy2014}.
In the notation of Tolstoy~\cite{Tolstoy2014}, the parastatistics algebra with relative paraboson relations would be $\mathfrak{osp}(1,2m|2n,0)$.
In~\cite{SV2018}, this $\Z_2\times\Z_2$-graded Lie superalgebra was denoted as $\pso(2m+1|2n)$.

The Fock spaces of order $p$ for the parastatistics algebra with relative parafermion relations were studied in~\cite{SV2015}.
They correspond to a class of lowest weight representations of the Lie superalgebra $\osp(2m+1|2n)$.
In a similar way, the Fock spaces of order $p$ for the parastatistics algebra with relative paraboson relations were studied in~\cite{SV2018}, 
corresponding to a class of lowest weight representations of the $\Z_2\times\Z_2$-graded Lie superalgebra $\pso(2m+1|2n)$.
Although one is dealing with different algebraic structures (in terms of gradings, commutators and anticommutators), the similarity between these representations was striking.
Now that we have identified the Klein transformation relating these structures, the similarity becomes completely clear.

\begin{acknowledgement}
N.I.\ Stoilova was supported by the Bulgarian National Science Fund, grant KP-06-N28/6, and J.\ Van der Jeugt was 
supported by the EOS Research Project 30889451. 
\end{acknowledgement}
\end{document}